# Towards Constraint-based High Performance Cloud System in the Process of Cloud Computing Adoption in an Organization

**Mikael Fernandus Simalango**
WISE Research Lab, Ajou University
Woncheon-dong Yeongtong-gu, Suwon, South Korea
mikael@ajou.ac.kr

**Mun-young Kang**
Department of Computer Engineering, Graduate School of Ajou University
Woncheon-dong Yeongtong-gu, Suwon, South Korea
hanamy@ajou.ac.kr

**Sangyoon Oh**
School of Information and Communication Engineering, Ajou University
Woncheon-dong Yeongtong-gu, Suwon, South Korea
syoh@ajou.ac.kr

**Abstract**: Cloud computing is penetrating into various domains and environments, from theoretical computer science to economy, from marketing hype to educational curriculum and from R&D lab to enterprise IT infrastructure. Yet, the currently developing state of cloud computing leaves several issues to address and also affects cloud computing adoption by organizations. In this paper, we explain how the transition into the cloud can occur in an organization and describe the mechanism for transforming legacy infrastructure into a virtual infrastructure-based cloud. We describe the state of the art of infrastructural cloud, which is essential in the decision making on cloud adoption, and highlight the challenges that can limit the scale and speed of the adoption. We then suggest a strategic framework for designing a high performance cloud system. This framework is applicable when transformation cloud-based deployment model collides with some constraints. We give an example of the implementation of the framework in a design of a budget-constrained high availability cloud system.

## 1. Introduction

Rising popularity of cloud computing not only in the IT sector but also in academia can indicate that the term is more than just a buzzword. Taking the recent trend in the industry, for example, steadily growing exposure by the media on cloud computing has incited further studies on the feasibility of its implementation in the enterprises or other commercial entities at smaller scale. Early cloud adopters, the technology companies with highly-adept IT force such as Google, Yahoo, Amazon, IBM and VMWare, have also been directly and indirectly involved in evangelizing this computing paradigm to wider audience in the industry and beyond.

Observing how cloud computing was introduced and popularized, it can be noticed that the main drives to the growing popularity of this paradigm come from enterprises. From an academic point of view, this is an interesting observation of technology development since the notion of an emerging mass technology is when it was initially invented in a research lab or academic institution, tested and iterated in a close community sponsored by the industry and finally introduced to public as a ready-to-use technology. The way we see the cloud appear from the horizon until we get a clearer sight of its shape is arguably the backward way; the academia uses and adopts a set of enterprise technologies, analyzes the eminent issues to tackle, and later overcomes those issues by developing a more integrated and generic solution after the core concept in the base technologies.

Following the observation, we notice that the enterprise drives in the popularization of the cloud are somewhat excessive so that the answer to the question if it is now the prime time for the cloud in most organizations can be biased. By mentioning an organization, we refer to an entity consisting of at least ten members and use computing technology in day-to-day operation to help achieve its goal. The minimum membership threshold, however, can be more or less as long as there is a defined structure or organizational hierarchy with minimum conflict of role and responsibility. We then argue that adopting cloud computing in an organization should not be a decision made as the reaction to a popular trend. We necessitate a due assessment conducted by the organization before deciding that a cloud-based system is the right architecture to answer the organizational needs.

The implication is hence two fold. Firstly, the organization should be well-informed about the state of

"This research was supported by the MKE(The Ministry of Knowledge Economy), Korea, under the ITRC(Information Technology Research Center) support program supervised by the NIPA(National IT Industry Promotion Agency" (NIPA-2010-(C1090-1021-0011)) cloud maturity. Secondly, it is also necessary to have a framework to validate the marginal benefit brought by the transformation from a legacy system into the cloud. To address the first implication, we describe the current state of infrastructural cloud that plays a big role in the transformation of physical infrastructure into a virtual infrastructure-based cloud system. Subsequently, we provide a case of an organization planning to transform its IT infrastructure into a cloud-based system. We explain possible approaches to the transformation and also show the technical artifacts of the infrastructural transformation.

We argue that even though utility-based pricing model and SLA guarantee offered by a public cloud provider is attractive, some concerns such as economic feasibility, performance stability, data integrity, security, SLA compliance and also organizational changes can be raised during the transition period to the virtual infrastructure-based cloud system. Special attention is put on the performance aspect of the cloud. In a retrospect, the shift into a cloud-based system should result in at least comparable performance with the legacy system. Earlier, some have speculated about the cloudy future of high performance computing [1][2]. In a similar nuance, we also argue that expecting high-performance in the cloud is bound by certain constraints. These constraints are set by the inherent factors contributing to performance penalty such as virtualization and the way a cloud is formed and consumed by an organization. Rephrasing these concerns as constraints, we propose a framework of strategies towards building a constraint-based high performance cloud system. To better exemplify the implementation of this framework, we show a design of budget-constrained high availability cloud system, which is modeled after the proposed framework.

The rest of the paper is organized as follows. In section 2, we discuss about the state-of-the-art of the infrastructural cloud. We review the cloud maturity and highlight several challenging questions yet to answer in managing the infrastructural cloud. Section 3 contains a discussion about the technical and managerial aspects of the process of transformation from a full-ownership, physical IT infrastructure into a partial-ownership, cloud-based IT infrastructure. In Section 4, we dive into more details about a strategic framework that can be applied by an organization planning to adopt a cloud-based provisioning model to support its day-to-day operation. Since we are interested in the high performance aspect, the framework is primarily tailored to assist in designing a high performance cloud system. We then exemplify the implementation of the proposed framework in a design of a budget-constrained high availability cloud system, which is presented in Section 5. In section 6, we discuss the related work in cloud computing adoption. To conclude our work, we summarize the discussion and our proposal in Section 7 and also envision the future research agenda.

## 2. State of The Art of Infrastructural Cloud

According to Google Trend [3], the beginning of public interests in cloud computing can be traced back to 2007. By reflecting to the timeline, the cloud is perceivably nascent and yet to become mature. Nevertheless, such enormous drives from big enterprises, the early cloud adopters, and thriving interests in the academia and industries have been contributing to accelerated development and widespread of the emerging paradigm. In this section, we will discuss about the states of key technologies in infrastructural cloud.

### 2.1 Virtualization

Virtualization is the key enabler to cloud computing. It manages the abstraction of physical resources as virtual machines (VMs). The virtual machines can be instantiated, terminated, migrated, stored and released thus adding elastic resource provisioning capability, which is one of the cloud characteristics. Major issues in virtualization technology are related with the performance metrics such as virtualization overhead [4][5][6][7], scalability [8][9], and effective allocation and distribution of virtual machines [10][11][12].

At the functional level, virtualization can be considered a mature technology. The hypervisor technology, for example, has been available widely as open source software such as Xen [13] and KVM [14] or proprietary software such as VMWare eSXi [15] and Oracle VirtualBox [16]. Depending on the existing physical infrastructure and the virtualization plan, an organization can choose the desirable virtualization solution to be implemented in order to set up a private cloud. A private cloud is defined as a cloud built by virtualizing existing internal infrastructure owned by the organization. Yet, it can be expanded through the integration with external infrastructural cloud service from the public cloud provider whose offer matches the stringent SLA required by the organization. This expansion creates a hybrid pool of cloud infrastructure in a private cloud so that this specific private cloud is also referred to as private/hybrid cloud [17].

### 2.2 Infrastructural Cloud Management

There are two perspectives of management of an infrastructural cloud: management from cloud owner's perspective and management from cloud tenant's perspective. Since our focus is on the IT infrastructure transformation in an organization, we will put more emphasis on cloud management from cloud owner's perspective. The tasks encompassed by the term cloud management can be categorized into two: engineering tasks and managerial tasks. We defer the discussion about the managerial tasks in Section 3 and elaborate only the discussion about the engineering tasks in this section.

In an organization, there can be various applications running. Each application may require different environment settings for its installation and execution. Additionally, an application can be used by different types and number of users. Since each user may invoke various functions supported by the application he/she is using, the workload of each application and overall applications will be dynamics. The dynamicity sometimes produces issues such as increased latency, temporary application unavailability, application crashes or failure, disk failure and so forth. In the legacy, physical infrastructure-based system, the resolution for these problems can escalate into a higher scale. As for example, an application failure may require a node reboot instead of merely restarting the application. This in turn may affect other applications running in the same physical node. Virtualization features process, memory, I/O, and network isolation for each of running virtual machines [18]. Root-cause analysis and formulation of solution for majority of the incidents listed earlier can then be localized at the level of VM instance. The engineering tasks are hence primarily related with managing and handling various issues of virtual machines.

The basic engineering task is in managing the instantiation or launching of VMs. A VM launch is managed by the hypervisor and controlled by automation script or manual supervision. In a bare virtualized system, VM should be manually instantiated by the site administrator. The administrator will use an existing OS image stored in the disk or removable media to be deployed as a VM by executing the virtualization commands provided by the hypervisor. Following the OS installation, the administrator will install the selected applications and verify their functionalities. This process is iterated until all applications are hosted in their respective VMs.

With the development of VM image management such as OpenNebula [19], Eucalyptus [20] and VMWare vCloud director [21], some degree of automation is added. The VM image management adds capability to store the preconfigured OS and applications bundled together with virtual resource specifications into VM image repository. To instantiate a VM, a VM image will be supplied to the virtual infrastructure manager that will read the resource specifications contained in the VM image. This capability reaps some benefits compared to the manual VM image deployment. The first benefit is shortening the time to install and test applications on top of the OS platform. Another benefit is quicker VM image deployment in case of replicating a VM image in order to provide load balancing capability or increase high availability. The third benefit is more immediate fault recovery in case of VM failure. A new instance of VM can be immediately taken from the repository and launched immediately. In general, these benefits improve the resilience of the system to the variability of the application workloads.

At the current state, existing cloud management solutions have yet to support fully-automated resource provisioning. A system architect should formulate and implement a policy to marshal the cloud management tasks. In a practical domain, this translates as these administrative activities: 1) monitoring the states of running VM instances, 2) defining and executing the schemes of VM replication and load balancing to maintain the high performance, 3) consolidating VM distribution and node utilization across the datacenter(s) and 4) executing fault tolerance and disaster recovery strategy to prevent potential loss caused by service unavailability. This situation underpins our effort in proposing a strategic framework that guides the process of adopting the infrastructural cloud model in an organization.

## 2.3 Cloud Federation

Cloud federation is defined as interconnecting two or more clouds to enable more scalable and elastic cloud service provisioning. A federated cloud is also dubbed as cloud of clouds or intercloud [22]. Such federation enables service provisioning to users by a pool of cloud providers including the one where the user is subscribed. Earlier, we have mentioned about private/hybrid cloud that combines virtualized internal resources with external resources from public cloud service provider. This type of cloud is principally a restrained federated cloud since 1) the federation is limited by the interface to public cloud providers supported by the cloud infrastructure manager and 2) the federation is not reciprocal in which the need for offloading workloads to another cloud comes from the private cloud and not vice versa.

Figure 1 illustrates our view of a federated cloud. In the figure, we exemplify three compute clouds namely cloud A, cloud B and cloud C as the members of a cloud federation. Each cloud is independent in the sense that it has the capability to provision compute or application service requests from its user base by utilizing its virtual infrastructure. By mentioning user, we refer to cloud application user or SaaS user who accesses various applications running on the cloud. At least one cloud frontend gateway is situated in each cloud. This frontend gateway acts as the contact point for the user request. It manages how the request is distributed to the appropriate provisioning virtual node. In the case of a request for data-intensive processing, the gateway can assume the request as a job and adopt one of several job distribution models that include batch processing (Condor [23] as an example), MapReduce and Split and Merge [24] to carry out the data processing.

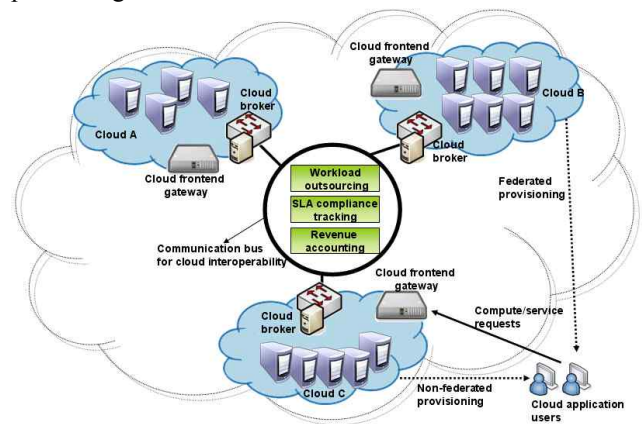

*Figure 1. A view of a federated cloud*

In anticipated workload conditions, a cloud is unconditionally self-sufficient, which means that it can provision its users' requests directly from within the cloud. Yet, variability of the workloads impacts the degree of the self-sufficiency and quality of service (QoS) delivered to the users. Degraded performance and overall QoS may result in SLA violation, a situation that is avoided by the cloud provider. For a public cloud provider, SLA violation may induce varying degrees of consequence ranging from the modest one, for example, new support ticket, to a more serious one such as contract termination by the tenant that incurs revenue loss. In a private cloud, the consequence is mostly related with organizational agility. If the organization is running on a cloud, frequent SLA violations can impact the duration for task execution and accomplishment, which decelerates the speed of organizational operations, and thus reducing its agility.

In Figure 1, we depicted a communication bus in a ring representation that interconnects the three independent clouds and thus forming a federated cloud. The communication bus enables each cloud to share its state and events it wants to multicast or broadcast over the bus. The events can be requests to outsource some workloads to another federated cloud member, updates of its status of outsourcing capability (supported VM image specs, supported SLA, contract duration, pricing) and workload execution status, and other types of notification messages (outsourcing offer, outsourcing confirmation, contract initiation, contract end, outsourcing fee confirmation, etc). Since it is important to track the state changes in the bus, a cloud broker can be placed at the edge of the cloud to manage and monitor the workload outsourcing, SLA compliance tracking, and revenue accounting. More detailed architecture of the cloud broker is beyond the scope of this article.

Cloud federation is a viable option for an organization with global reach in terms of users or office representations. We will use an example of an organization with global office representations. Let us mention a global IT company X with offices located in the continents of Asia, Europe, and America. Assuming that the company has built one datacenter for every continent, after the transition of the datacenter into infrastructural cloud model, it will practically own three private clouds. Each private cloud can be of different capacity. The workload and utilization level are also varying by time depending on the types of applications running and the consumers of each datacenter. Federating these clouds enables cross sharing of virtual resources, saving some unnecessary cost to provision extra physical infrastructure, and also reducing energy usage through the consolidation of the datacenters.

Cloud federation is a moving target and currently not an established technology notwithstanding the potential benefit it carries. There are currently several issues yet to answer. The first issue is a common communication protocol for a federated cloud. The lack of standard sends the efforts of federating the clouds to different avenues. OpenNebula and Eucalyptus, for example, uses a driver-based approach to integrate a private cloud with other infrastructural cloud and increase the elasticity. While this approach is reasonable, the apparent issue is in the complexity of creating and updating the drivers for the clouds that join the federation. The next issue is limited degree of the automation of workload outsourcing process. There has not been a widely accepted cloud federation model that harnesses automatic workload outsourcing and its relevant cloud economy. We notice that there have been efforts towards a more standardized federated cloud, for example, the open federated cloud model RESERVOIR [25] and Intercloud Protocols proposal [26] from Cisco. Yet, such effort is still work on progress and has not been validated in real case of federated cloud. Several other issues notably security and trust, naming and addressing, and data management are impending thus requiring further investigation to get resolved.

## 3. Towards Transformation into Cloud-based IT Infrastructure

We have extensively discussed about the current state-of-the-art of the cloud computing, primarily from the outlook of an infrastructural cloud. The notion that cloud is disruptive in terms of the paradigm shift it brings, which is on-demand utility-based internet computing, is potential to bring changes to the way an organization plan and manage its accounting and IT infrastructure. In a contemporary perspective of IT, computer hardware, software and telecommunications equipment are defined as IT assets [27][28]. Among the IT assets, the idea and practice of utility-based software or software as a service (SaaS) came earlier before utility-based infrastructure was introduced as a part of the cloud services. Hence, we are interested in discussing the shift in perspective about equipment and hardware, the physical infrastructure, which is regarded as fixed assets in contemporary accounting.

Traditionally, the physical infrastructure is purchased by the organization and it then owns sole ownership to the infrastructure. Nevertheless, its economic valuation will be depreciated annually for as long as its useful life until the valuation reaches a minimum value, which then promotes a hardware or equipment replacement. Purchasing the infrastructure requires significant investment or capital expenditure (CAPEX) [28] notwithstanding the decreasing trend of hardware prices [27]. Yet, such significant upfront investment does not immediately translate as the maximum utilization of the hardware and equipment. Servers in a datacenter are typically running at the levels that are below the safe maximum capacity [29]. Another more obvious example is storage devices, which are initially at the lowest utilization and gradually filled up to its maximum capacity.

Comparing the economic aspect of traditional IT infrastructure provisioning with utility-based infrastructure provisioning ala public cloud can encourage the stakeholder to plan for cloud adoption across the organization. Smaller upfront payment, flexible contract, immediate scaling made possible by renting public cloud services are some incentives that can motivate the cloud adoption. In an enterprise perspective, these incentives are translated as the prospect of higher return of investment (ROI). Besides the economic benefit, an organization that transforms its existing infrastructure into a cloud can increase the utilization of its servers and also reduce the number of servers used by consolidating the placement of VMs. Consequently, this effort will reduce the consumption of energy and indirectly support the green IT campaign.

Despite the benefits that cloud computing offers, there are several challenges that can affect the decision of cloud computing adoption. The low cost of scaling, for example, does not directly correlate with performance and QoS that can be assured by a public cloud. Some studies about the performance of EC2 cloud [30][31][4][32] revealed degradation of the performance metrics such as average latency and network throughput, and increased variability of the observed metrics compared to their behavior in native, non-virtualized system. Aside from this performance problem, there are also fundamental issues that the cloud introduces. In a private cloud, the issues include various aspects of virtualization and management of the cloud. In a multi-tenant cloud, additional issues such as privacy and confidentiality, security and trust, network addressing, and service compliance also exist. Having these issues are undesirable especially for the enterprises since it can affect its image and business continuity.

There is a gap between economic incentives primarily offered by the cloud with the inflected changes it undertakes in an organization. We argue that the changes will be dispersed to the entire organization instead of only at the internal scale of the organization's IT force. Changes at the management level, for example, include varying degree of reforms in the accounting and bookkeeping system, IT-related task delegation and evaluation, infrastructural planning, and strategy for organizational growth. The IT force as the main caretaker of the infrastructural transformation will oversee the process of asserting, planning, executing and evaluating the virtualization of current infrastructure and the integration with public cloud infrastructure based on the profiles of applications used by the organization. The changes in this unit are mostly related with the changes of responsibility, which is from the main executor for the planning, setup, and maintenance of the physical infrastructure into a consultant or certifier of the cloud on which the organization is running. Other members of the organization can experience the behavioral changes of how they use an application, save and retrieve data, and get the resolution for issues or incidents related with the application they are using. As an example, cloud application users generally do not have to install the application on their PC. Since the application is running on the cloud, they use the interface provided (e.g. web interface) to launch the application and then save their work primarily to the cloud storage instead of the local hard drive. Additionally, problem with the application they are using may not be resolved immediately if it is caused by glitch or failure at cloud provider side.

To put the discussion about the process of cloud computing adoption in a summary, we argue that the adoption will be gradual as the organization gets better understanding about cloud computing and cloud computing technologies become more mature. We also envision that the adoption process will require a trial period where certain legacy application is migrated into the cloud and its functionalities are tested until the organization is ready to run on the cloud.

## 3.1 A Perspective of Approaches to Cloud Computing Adoption in an Organization

We argue that the cloud adoption in an organization can proceed with two approaches: top-down approach and hybrid approach. In top-down approach, the initiative for the adoption comes from the stakeholders who define the transition into cloud as a part of the organizational strategy. The IT force will plan the adoption by analyzing current infrastructure and profiling the applications used throughout the organization. Following the analysis, the unit asserts the adoption and then executes the transition into a cloud-based infrastructural model. After successfully checking the functionalities of the affected applications, the transformation is pushed to users who will then start running the applications in the cloud.

The hybrid approach, in contrast, starts the cloud adoption as an internal process in the IT force. The unit initiates limited transformation into the cloud through the migration of a test application into a cloud testbed. The migration will be followed with testing by cloud application users. Tickets containing functional defect, feedback, and other requests are sent back to the IT force during the testing phase. The tickets will be investigated and resolved in order to ensure the readiness of organization-wide adoption. This process is iterated until reaching the desired state, which can be quantified as the SLA compliance with a set of parameters. The state of cloud adoption readiness will be promoted to the stakeholders who will decide if the transformation will be applied to the whole organization. The comparison of process flows in both approaches can be seen in Figure 2.

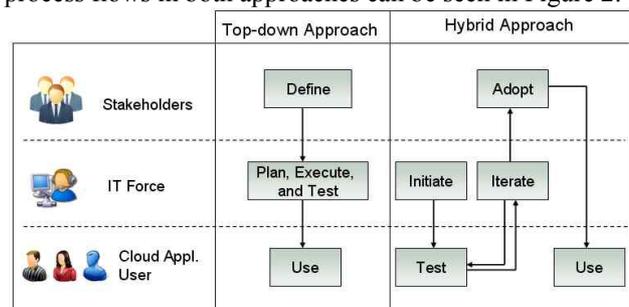

*Figure 2. Comparison of approaches to cloud adoption in an organization*

In our opinion, hybrid approach is more potential to bring success to the cloud adoption in an organization. This is because of there is sufficient iteration process for testing and evaluation that is conducted before the cloud deployment model is applied to the entire organization.

## 3.2 The Technical Artifacts of Transformation into Cloud-based IT Infrastructure

Regardless of the cloud adoption approaches, we find out that there is a common technical procedure that can be carried out in the transformation of legacy IT infrastructure into a virtual infrastructure-based cloud. The procedure is applicable at datacenter-wide. We assume that applications are detachable from their storage or data management system so that we can bundle the applications independently of their storage. In the real world case, the procedure requires the datacenter to have NAS (Network Attached Storage) or SAN (Storage Access Network) capability.

We formalize the procedure as the following. There are $N$ nodes running as servers in the datacenter C, hence $C = \{node_1, node_2, ..., node_n, ..., node_N\}$. We obtain the list of the applications $S = \{s_1, s_2, ..., s_m, ..., s_M\}$ running on the physical infrastructure and get $M$ numbers of applications and $K$ number of operating systems. We then denote the collection of the operating systems as $OS = \{os_1, os_2, ..., os_k, ..., os_K\}$. Since there can be several applications in one operating system, it is obvious that $k \leq K \leq M$. Suppose that each application needs storage to keep the intermediary data or processed output. We assume $G$ number of storage nodes, $G < N$, and assign storage space for each application. By denoting the index of the storage as $g \in [1, N]$, the list of storage space for all applications can be denoted as $D = \{d_{1, g \in [1,N]}, d_{2, g \in [1,N]}, ..., d_{m, g \in [1,N]}, ..., d_{M, g \in [1,N]}\}$.
To start the virtualization, we define a set of VM image templates. Each template contains OS, CPU, memory, and disk specification. Let us assume that there are $H$ types of templates so that $VMI = \{vmi_1, vmi_2, ..., vmi_h, ..., vmi_H\}$ and $vmi_h = <OS_h, CPU_h, mem_h, disk_h>$.

In a case of private cloud, the rest of the transformation process is listed in order as the following:
1. We classify the applications according to their operating systems. After the classification we will have as many as $K$ classes of application-OS pairs.
2. For each application-OS pairs in a class, we refine the applications that we want to run in the same VM and the one that we will put into separate VM. We repeat this step for the other classes. At the end of the process, we will have $y \leq M$ VMs to instantiate.
3. In case of the availability of VM image creation utility, we can build the images for the OS-application pairs that are obtained from 2. The images can be stored in node $g \in [1, N]$ or another node in C.
4. We plan the instantiations of $y$ number of VMs over the ($N - G$) nodes based on installed capacity and existing utilization of each node.
5. We execute the instantiation of VMs on all nodes counted as the target of virtualization obtained in 4.
6. For each instantiated VM, we configure it to reconnect the application(s) contained with previously designated storage space(s).

Summarizing the technical artifacts of the infrastructural transformation, we can conclude that the base process is primarily adding virtualization layer on existing infrastructure and consolidating the servers in the datacenter by managing the distribution of the VMs across the datacenter. Big datacenter can incorporate additional stages such as adding VM management layer, building an enhanced VM image repository for quicker VM deployment or other automation process.

## 4. A Strategic Framework for Constraint-based High Performance Cloud System

The procedure of cloud computing adoption we discussed earlier uses an optimistic approach to transforming the infrastructure and migrating applications into the cloud. In reality, the transformation will not be as straightforward as explained. Let us use an example of application profile diversity. Based on its functionality, we can categorize an application used in an organization into two: mission critical and non-mission critical. A mission critical application requires always-on, minimum failure, low latency, high throughput, high reliability, and other characteristics indicating high performance. This strict requirement entails different strategy in managing the application in the cloud – if cloud is used as the deployment model. To ensure the application is always-on or highly available, the application can be replicated to provide redundancy and at the same time load balancing capability. To maintain low latency, the application should be loaded from the closest network distance against user's location. Additionally, the application workload in the VM running the application should also be monitored so that it does not suffer from resource drain that can result in performance degradation.

### 4.1 Constraints in Designing a Cloud System and the Argumentation

We state three prepositions that converge into an idea that a design of a high performance cloud system is bounded by some constraints. Our propositions are explained as follows.

**Proposition 1** In a private cloud, number of replication is bounded by the infrastructural size
**Argument:**
We will certify this proposition by a contradiction. In Section 3.1, we have $N$ nodes and $M$ applications. We have $M$ virtual machines to be instantiated in case of one-to-one mapping of application-OS and $y < M$ virtual machines in case of many-to-one mapping. Suppose all nodes have the same capacity and all VMs uses the same resource specification $VM_{spec}$. Define $k$ as the maximum number of VMs with $VM_{spec}$ that can be launched by a node hence we have $Nk$ VMs that can be launched by the datacenter. By distributing the VMs evenly to all nodes, there will be $\frac{y}{N} < k$ or $\frac{M}{N} \leq k$ VMs in each node. For the critical case $M = Nk$ and no VM release, adding another replica of VM will change the total launched VMs into $M' = M + 1 > Nk$, which is contradictive with the maximum $Nk$ VMs that can be launched.

**Proposition 2** Replicating an application in the public cloud does not always improve the performance
**Argument:**
We draw an overlay network that illustrates a cloud application user, a private cloud, and a public cloud in Figure 3. The private cloud is denoted as cloud A while cloud B denotes the public cloud. A user is subscribed to cloud A and runs an application placed in this cloud. A restrained federation is enabled between cloud A and cloud B hence cloud A can outsource some workloads to cloud B.

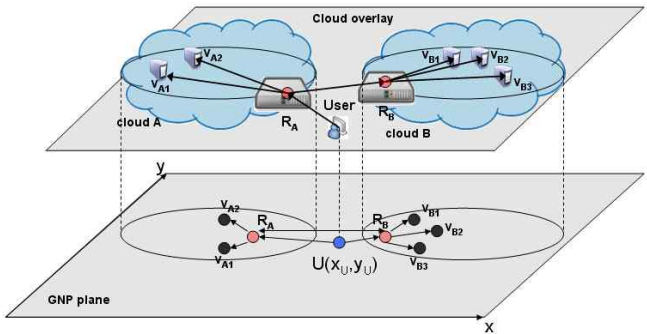

*Figure 3. Clouds and a cloud application user represented in a GNP plane*

Application performance can be measured by several metrics such as latency, network throughput, jitter, bit rate, and so forth. We choose latency and use network distance as the criteria to measure this metric. Assume that all nodes in the cloud A and cloud B will launch VM of the same specification so that we can eliminate the factor of processing capability. We also assume that all VM will be assigned a public IP address so that they are reachable from global network. Suppose that we have obtained the network coordinates [33] of the client and other nodes. We then create a Global Network Positioning (GNP) plane to map these coordinates. Let us define $d_{E-F}$ as the network distance from node E to node F and $\omega_{E-F}$ as the network speed between node E and node F.

We will provide a case of load balancing an application. In this case, an application is accessible in both private cloud A and public cloud B. To reduce the complexity of analysis, let us assume that VM located in the physical node can directly provision the request to user. We can compute the latency as follows. First, we measure the time to provision the request $t_{prov}$, which is obtained by adding the time for the request to reach cloud frontend gateway $t_{U-R}$ with processing time at gateway $t_R$, the time for dispatching request to a VM $t_{R-v}$, processing time at virtual machine $t_v$, and response delivery time from VM to user $t_{v-U}$. We then compare the total provisioning time of cloud A $t^A_{prov}$ with provisioning time of cloud B $t^B_{prov}$.

$$t^A_{prov} = t^A_{U-R} + t^A_R + t^A_{R-v} + t^A_v + t^A_{v-U}$$

$$t^A_{prov} = \frac{d^A_{U-R}}{\omega^A_{U-R}} + t^A_R + \frac{d^A_{R-v}}{\omega^A_{R-v}} + t^A_v + \frac{d^A_{v-U}}{\omega^A_{v-U}} \quad (1)$$

$$t^B_{prov} = \frac{d^A_{U-R}}{\omega^A_{U-R}} + t^A_R + \frac{d^{fed}_{A-B}}{\omega^{fed}_{A-B}} + t^B_R$$
$$+ \frac{d^B_{R-v}}{\omega^B_{R-v}} + t^B_v + \frac{d^B_{v-U}}{\omega^B_{v-U}} \quad (2)$$

$$t^B_{prov} = \frac{d^B_{U-R}}{\omega^B_{U-R}} + t^B_R + \frac{d^B_{R-v}}{\omega^B_{R-v}} + t^B_v + \frac{d^B_{v-U}}{\omega^B_{v-U}} \quad (3)$$

Based on the equations above, Eq. (2) is applicable when the user can not request directly from cloud B and should let the frontend gateway of cloud A to reroute the request. In brief, the equation is applicable in the case of a restrained cloud federation. In Eq. (3), such limitation is removed thus enabling direct access from cloud B. This equation is applicable in the case of a full cloud federation. We are interested in obtaining the condition that satisfies $t^B_{prov} < t^A_{prov}$. Since we have the same VM specification, we can consider $t^A_R \approx t^B_R$ and $t^A_v \approx t^B_v$. If we assume that the gateway can find the VM located very close to it, we can consider $d^A_{R-v} \approx d^B_{R-v}$. Subsequently, we can conclude that provisioning the request from a public cloud B can improve the performance if 1) a full federation is involved, 2) the user is located closer to the public cloud in the GNP plane and 3) the network speed from user to this cloud and inside the public cloud is faster or at least comparable with that of the private cloud.

**Proposition 3** Given a budget cap, application performance can be sustained if there exists a workload pattern.
**Argument:**
To save some space, we explain the reasoning for this proposition briefly without mathematical expression. Referring back to Figure 3, we now add a cost function on cloud A and cloud B. We have a specific budget cap for running application X in the cloud. However, we assume that provisioning from internal resources is not affected by the budget cap. Suppose that we have obtained the profile of application X workload. If there is a workload pattern, we can split the pattern into several slices. For the slice that contains high workload, we outsource some workload to a public cloud whose offer match the SLA desired for the application. For the slice with low workload, we try to minimize extra expense by preventing workload outsourcing. By using this approach, we can project better to stay under the budget cap while maintaining the compliance with the SLA.

## 4.2 A Framework for Designing Constraint-based High Performance Cloud System

Since designing a cloud system is not purely a matter of technical sophistication, we necessitate the incorporation of a strategic framework into the process of system design. The strategic framework should include various aspects that should be considered in designing a cloud system. In Figure 4, we show the process flow of the strategic framework.

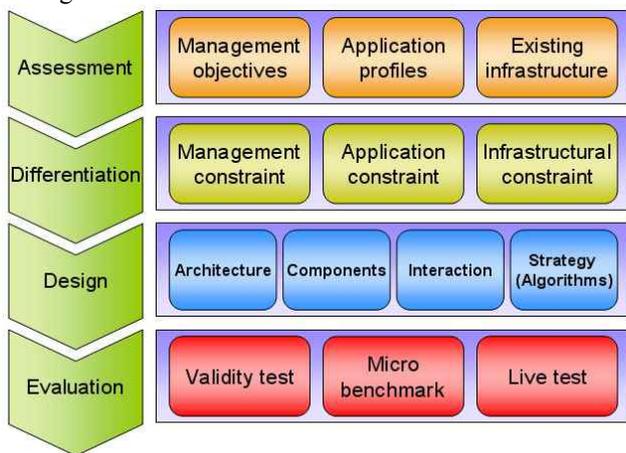

*Figure 4. Our suggested framework for constraint-based cloud system design*

Our proposed framework consists of four stages namely assessment, differentiation, design and evaluation. In each stage, there are several coarse-grained sub-processes that can be customized by the framework implementer. Assessment stage aims at gathering checklists about the state of readiness of an organization in adopting cloud computing. This stage comprises the sub-processes that collect the information about management objectives, application profile, and existing infrastructure. A framework implementer should define a set of checklists that help reduce the constraints of the system design. As for example, the checklist in application profile can include application support to parallelism, decoupling of application from storage, application reliability, etc while checklist in existing infrastructure can include the availability of local cluster as a private cloud and cloud management solution for elastic scaling.

Differentiation stage separates the verifiable checklists with non-verifiable ones. Non-verifiable checklists will be used as a constraint that differentiates the design of the cloud. Since we are interested in the high performance aspect, the differentiation strategy is centralized around maintaining the high performance. In the design stage, this is translated as a good architectural design, a pool of provisioning strategy, proper component selection and effective interaction among components to reach and maintain the target performance by taking care of the constraints. The final stage, evaluation, provides several methods to assert the acceptability of the design. In the most basic test, the design will be fed into a simulation engine. This engine will then generate various scenarios for application provisioning as a validity test. If the design exhibits compliance to the performance requirement, a more thorough micro benchmark and live test can be conducted to ensure the resilience of the design with more varying workload patterns.

## 5. Design Case Study: A Budget-constrained High Availability Cloud System

In this section we show how we formulate the design of a cloud system by referring to our proposed framework. We wanted to design a high availability cloud system with budget as the main constraint. By following the process flow described in our framework, we start from the assessment stage by gathering the management objectives, application profiles, and size of existing infrastructure. The objective of the management is to maintain a range of budget for the cloud deployment. As for the application profiles, there are several applications that should be migrated into the cloud and these applications can be packaged into a VM image without much dependency. On the other hand, existing infrastructure, which supports SAN storage, has been virtualized to form a private cloud.

We then move to the second stage, which is differentiation. Based on the assessment in the first stage, the primary constraint of the design is budget. The constraint of the application is only its number. As for the infrastructure, the readiness of the private cloud literally means that we can use the internal resources and then scale out to public cloud when necessary.

The differentiation results in the design of such cloud system illustrated in Figure 5. In the figure, we can see the architecture, components and interaction among them.

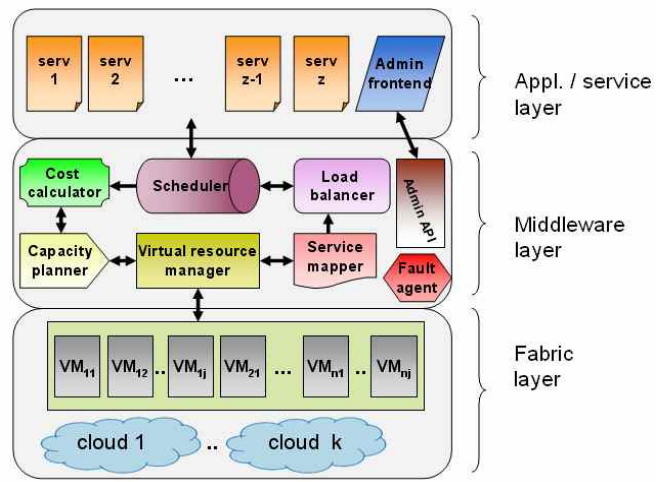

*Figure 5. A design of high availability cloud system with budget as a constraint*

We need a virtual resource manager since we will scale out to public cloud. We also need a cost calculator to calculate and update current billing state to the scheduler. The scheduler is the main actor that executes the strategy towards maintaining the high availability of the system.

Given that the system should support a variety of applications, we need a service mapper component, which will provide the information about the location of VMs and the applications tied with them.

In another design where budget is not a constraint, we can remove the cost calculator component from the design and instead focus on optimizing the scheduler to use more relaxed load balancing policy by instructing capacity planner to create more replicas and then execute load balancing policy on a bigger pool of VMs.

## 6. Related Work

In the journey to wider adoption of cloud computing, we have observed how the early cloud adopters leverage cloud computing in their organization. Additionally, we include another case study and review other models proposed to be implemented in the process of cloud computing adoption.

### 6.1 Cloud Computing Adoption by Enterprise Forerunners

Google introduced their MapReduce programming model [34], which is applicable for data-intensive compute task processing in the cloud. It also offers a public use of its Google App Engine [35], a type of Platform as a Service (PaaS) that can be used as a platform to build Python or Java-based web application on top of Google infrastructure. Yahoo has been actively involved in testing and developing Hadoop [36], a platform that can be used to handle various distributed computing and data management tasks including the implementation of MapReduce programming model. Besides, it has also introduced PNUTS [37], a distributed database system that applies publish/subscribe paradigm, for running Yahoo!'s web applications.

Amazon introduced its Amazon EC2 [38], a type of Infrastructure as a Service (IaaS) cloud service that has gained popularity in the testing and benchmarking of various aspects of infrastructure cloud such as performance [30][31][4][2], economies of scale [39][40][41] and security [42][43][44]. IBM has also been intensively researching on the cloud and disseminating their work to the scientific peers. Some groundbreaking work includes Blue Eyes [45], a system management in the cloud, and IBM's global testbed for compute cloud, which is named after RC2 [46]. In addition to its cloud research initiatives, the company has also started to offer a range of cloud solutions to the enterprises [47]. VMWare has been focusing on developing the key enabler technology for cloud computing, which is virtualization. The company has been known for its virtualization solutions that help transform legacy, hardware-based enterprise IT infrastructure into virtual infrastructure-based cloud. Besides the examples of early cloud adopters specifically mentioned here, there are currently several other commercial entities developing various aspects of the cloud, which in turn help this emerging computing paradigm on the way towards its maturity.

### 6.2 Modeling and Case Study of Cloud Computing Adoption

There is still currently a little scholarly work that discusses about cloud adoption in organization or enterprise and its undertaking. A notable study was carried out by Hosseini *et al.* [48] who analyzed the impacts of adopting IaaS in an organization and the implications of the adoption for the cloud application users. They also suggest Cloud Adoption Toolkit [49], a toolkit that can be used in the process of decision making about cloud adoption. Chang *et al.* [50] reviewed cloud cube model (CCM), a cloud business model that can be adopted by the management of an organization, and suggested a hexagon model for sustainability of the adoption.

## 7. Conclusion and Future Work

Rapid proliferation of cloud computing in today's internet computing arena has incited the momentum of wider adoption of this paradigm in organizations, either IT-inclined or non IT-inclined. The adoption includes the transformation of physical IT infrastructure into virtual infrastructure-based cloud and integration with external cloud services to improve the elasticity. We showed the current state of the art of infrastructural cloud to give consideration for organizations planning to adopt the virtual infrastructure-based cloud model. The adoption process itself can be undertaken in two approaches but the adopter can observe a common technical procedure in transforming the infrastructure into an IaaS cloud.

We have shown that due to various organizational needs and the state of cloud maturity, a cloud system will be built on different criteria or constraints. This condition was the motivation behind the framework for the constrained-based high performance cloud system that we proposed. With its existence, we have experienced how a cloud system design can be developed to fit its constraints and focus on its performance target.

Our future work consists of validating the usability of the framework in general cases and providing the evaluation of cloud system designs derived from the process flow contained in the framework.